\documentclass{ifacconf}

\usepackage{graphicx}      
\usepackage{natbib}        
\usepackage{amsmath}
\usepackage{amssymb}
\usepackage{subcaption}
\usepackage{mathtools}
\usepackage{float}
\usepackage[algo2e, linesnumbered]{algorithm2e}
\usepackage{color}
\usepackage[normalem]{ulem}

\newcommand{\RR}{\mathbb{R}^{2}}
\newcommand{\Qi}{Q_{i}}
\newcommand{\Qj}{Q_{j}}
\newcommand{\sgnDist}{b}
\newcommand{\pix}{p_{i,x}}
\newcommand{\piy}{p_{i,y}}
\newcommand{\vix}{v_{i,x}}
\newcommand{\viy}{v_{i,y}}
\newcommand{\uix}{u_{i,x}}
\newcommand{\uiy}{u_{i,y}}
\newcommand{\pjx}{p_{j,x}}
\newcommand{\pjy}{p_{j,y}}
\newcommand{\vjx}{v_{j,x}}
\newcommand{\vjy}{v_{j,y}}
\newcommand{\ujx}{u_{j,x}}
\newcommand{\ujy}{u_{j,y}}

\newcommand{\vxr}{v_{r,x}}
\newcommand{\vyr}{v_{r,y}}
\newcommand{\pxr}{p_{r,x}}
\newcommand{\pyr}{p_{r,y}}
\newcommand{\vdxr}{\dot{v}_{r,x}}
\newcommand{\vdyr}{\dot{v}_{r,y}}
\newcommand{\pdxr}{\dot{p}_{r,x}}
\newcommand{\pdyr}{\dot{p}_{r,y}}

\newcommand{\ttr}{\psi}
\newcommand{\rd}{r_d}

\newcommand{\qedwhite}{\hfill \ensuremath{\Box}}
\let\oldnl\nl
\newcommand{\nonl}{\renewcommand{\nl}{\let\nl\oldnl}}

\begin{document}

\begin{frontmatter}

\title{Safe Coverage of Compact Domains For Second Order Dynamical Systems}


\author[First]{Juan Chacon} 
\author[Second]{Mo Chen} 
\author[Third]{Razvan C. Fetecau}

\address[First]{Department of Mathematics, Simon Fraser University, Burnaby BC, Canada, V5A 1S6, juan\_chacon\_leon@sfu.ca}
\address[Second]{School of Computing Science, Simon Fraser University, Burnaby BC, Canada, V5A 1S6, mochen@sfu.ca}
\address[Third]{Department of Mathematics, Simon Fraser University, Burnaby BC, Canada, V5A 1S6, van@sfu.ca}

\begin{abstract}                
Autonomous systems operating in close proximity with each other to cover a specified area has many potential applications, but to achieve effective coordination, two key challenges need to be addressed: coordination and safety.
For coordination, we propose a locally asymptotically stable distributed coverage controller for compact domains in the plane and homogeneous vehicles modeled with second order dynamics with bounded input forces. 
This control policy is based on artificial potentials designed to enforce desired vehicle-domain and inter-vehicle separations, and can be applied to arbitrary compact domains including non-convex ones.
We prove, using Lyapunov theory, that certain coverage configurations are locally asymptotically stable. For safety, we utilize Hamilton-Jacobi (HJ) reachability theory to guarantee pairwise collision avoidance. 
Rather than computing numerical solutions of the associated HJ partial differential equation as is typically done, we derive an analytical solution for our second-order vehicle model. 
This provides an exact, global solution rather than an approximate, local one within some computational domain.
In addition to considerably reducing collision count, the collision avoidance controller also reduces oscillatory behaviour of vehicles, helping the system reach steady state faster. 
We demonstrate our approach in three representative simulations involving a square domain, triangle domain, and a non-convex moving domain.
\end{abstract}


\end{frontmatter}

\section{Introduction}
\label{sect:intro}

Autonomous systems have great potential to positively impact society. 
However, these systems still largely operate in controlled environments in the absence of other agents.
Autonomous systems cooperating in close proximity with each other has the potential to improve efficiency.
Specifically, in this paper we consider the problem of controlling multiple autonomous systems to cover a desired area in a decentralized and safe manner.
To achieve effective coordination in this context, two key challenges need to be addressed: coordination and safety.

In coverage control problems, the objective is to deploy agents within a target domain such that they can achieve an optimal sensing of the domain of interest. A common approach to the coverage problem is by means of Voronoi diagrams (\cite{Cortes_etal2004,Gao_etal2008}), where the goal is to minimize a coverage functional that involves a tessellation of the domain and the locations of vehicles within the tesselation. This results in a high-dimensional optimization problem that has to be solved in real time.  In our approach we achieve coverage (alternative terminologies are balanced or anti-consensus configurations) through swarming by artificial potentials (see \cite{LeonardFiorelli2001,Sepulchre_etal2007} and also, the recent review by \cite{Chung_etal2018}). In a related problem, artificial potentials have been used for containment of follower agents within the convex hull of leaders (\cite{RenCao2011,Cao_etal2012}).

\begin{figure}
\centering
    \begin{subfigure}[b]{4.2cm}
        \includegraphics[width=4.2cm]{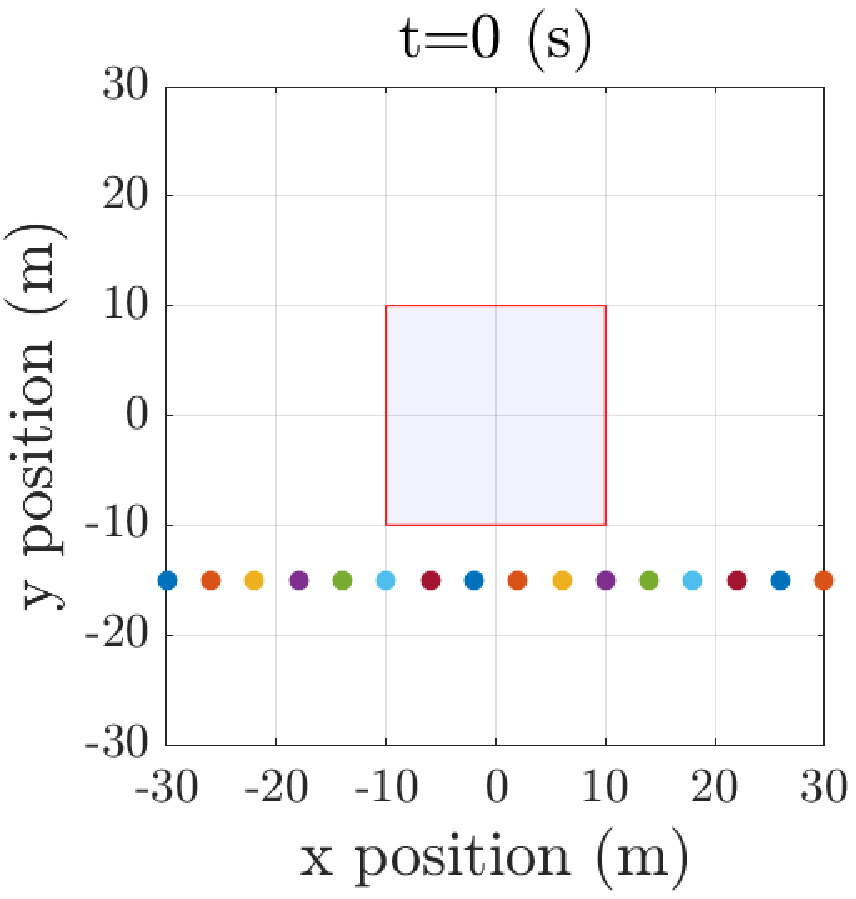}\hspace{-0.2cm}
        \caption{}
        \label{fig:squareInitial}
    \end{subfigure}
    \begin{subfigure}[b]{4.2cm}
        \includegraphics[width=4.2cm]{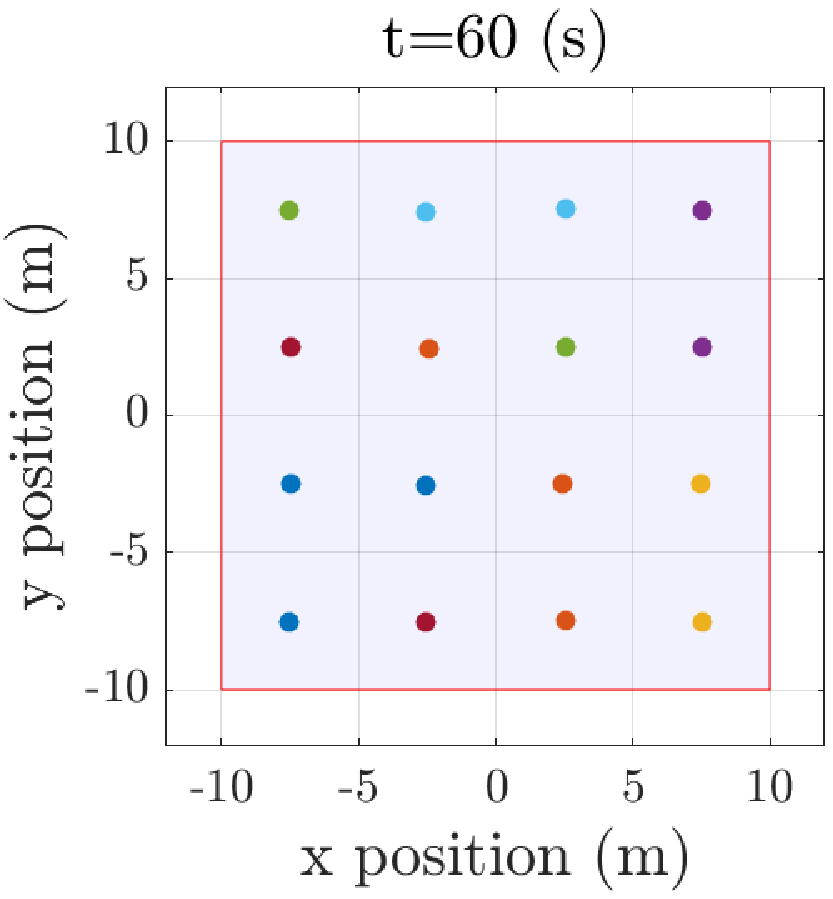}
        \caption{}
        \label{fig:squareSteady}
    \end{subfigure}
\caption{Initial (a) and steady (b) states for covering a square domain, with a square number of vehicles $\left(N=16\right)$.} 
\label{fig:InitialNSteadySquared}
\end{figure}

Reachability analysis as a safety verification tool has been studied extensively in the past several decades (\cite{Althoff2011, Frehse2011, Chen2013}).
In particular, Hamilton-Jacobi (HJ) reachability (\cite{Yang2013,Chen2018}) has seen success in applications such as collision avoidance (\cite{Alam11,Chen2016}), air traffic management (\cite{Margellos13,Chen2018a}), and forced landing (\cite{Akametalu2018}). 
HJ reachability analysis is based on dynamic programming, and involves solving an HJ partial differential equation (PDE) to compute a backward reachable set (BRS) representing the set of states from which danger is inevitable. 
Safety can therefore be guaranteed, despite the worst-case actions of another agent, by using the derived optimal controller when the system state is on the boundary of the BRS.

\textit{Contributions}: In this paper we develop a new approach to self-collective coordination of autonomous agents/vehicles that aim to reach and cover a target domain. The main factors that we consider in our approach are: i) reach  and spread over the target domain without having set \textit{a priori} the coverage configuration and the final state of each vehicle, ii) have a distributed control of vehicles in the absence of any leaders, that allows for self-organization and intelligence to emerge at the group level, and (iii) guarantee collision-avoidance throughout the coordination process.

In this aim, we consider a control system that includes both a coverage and a safety controller. The coverage controller is designed to bring the vehicles inside and spread them over the target domain, while the second guarantees collision avoidance of vehicles. In particular, the coverage controller uses artificial potentials for the inter-individual forces which are designed to achieve a certain desired inter-vehicle spacing (\cite{LeonardFiorelli2001}). Such controllers enable emergent self-collective behaviour of the vehicles, similar to the highly coordinated motions observed in biological groups (e.g., flocks of birds, schools of fish); see \cite{Camazine_etal2003}. 

We emphasize that the coverage controller proposed here (which also includes the approach to the target) is done through agent swarming; there is no leader and no order among the agents. This feature offers robustness to the controller, as it does not have to rely on the well functioning of each individual agent. Self-collective and cooperative behaviour in systems of interacting agents have been of central interest in physics and biology literature (see \cite{Vicsek_etal1995,Couzin_etal2002,DOrsogna_etal2006,CuckerSmale2007,FetecauGuo2012}). A collaborative robot search and target location (without coverage) based on a swarming model was done in \cite{Liu_etal2010}.

Our model is second-order, where agents are controlled through their acceleration; this is to be contrasted with first-order models, where the control inputs are the agents' velocities. We set \textit{a priori} bounds on the control forces, making our controller more realistic than previous approaches, where infinite forces were needed to guarantee collision avoidance (\cite{LeonardFiorelli2001,HusseinStipanovic2007}). 
For an illustration, we show in Fig. \ref{fig:InitialNSteadySquared} the initial and final states for a simulation using our controller for $N=16$ vehicles that cover a square domain.

The safety controller is derived from HJ reachability analysis. Unlike the typical approach of numerically solving an associated HJ PDE, we derive the analytical solution to the PDE to eliminate numerical errors and computation bounds. While multi-vehicle collision avoidance is in general intractable, we observe drastically reduced collision rate by just considering pairwise interactions.







\section{Background}
\subsection{Signed Distance}
\label{subsect:distance}
An oriented measure of how far a point $x$ is from a given domain $\Omega$ is the signed distance, defined by
\begin{align*}
\sgnDist\left(x\right)=\begin{cases}
\underset{x_{p}\in\partial\Omega}{\min}\left\Vert x-x_{p}\right\Vert , & \text{if}\,x\notin\Omega\\
-\underset{x_{p}\in\partial\Omega}{\min}\left\Vert x-x_{p}\right\Vert , & \text{if}\,x\in\Omega.
\end{cases}
\end{align*}

If $\nabla \sgnDist \left(x\right)$ exists, then there exists a unique $P_{\partial\Omega}\left(x\right)\in\partial\Omega$, called the projection of $x$ on $\partial\Omega$, such that
\begin{align}\label{eq:sgnDistChar}
    \sgnDist\left(x\right)=\begin{cases}
    \left\Vert P_{\partial\Omega}\left(x\right)-x\right\Vert , & \text{if} \,x\notin\Omega\\
    -\left\Vert P_{\partial\Omega}\left(x\right)-x\right\Vert , & \text{if} \,x\in\Omega
    \end{cases}
\end{align}
and
\begin{align}\label{eq:sgnDistGrad}
    \nabla \sgnDist \left(x\right)=\frac{x-P_{\partial\Omega}\left(x\right)}{b\left(x\right)}.
\end{align}


\subsection{Hamilton-Jacobi Rechability}
\label{subsect:HJ-reach}
Consider the two-player differential game described by the joint system
\begin{align}\label{eq:syst-gen}
\dot{z}\left(t\right) & =f\left(z\left(t\right),u\left(t\right),d\left(t\right)\right)\\
z\left(0\right) & =x,\nonumber
\end{align}

\noindent where $z\in\mathbb{R}^{n}$ is the joint state of the players, $u\in\mathcal{U}$ is the control input of Player 1 (hereafter referred to as ``control'') and $d\in\mathcal{D}$ is the control input of Player 2 (hereafter referred to as ``disturbance'') .

We assume $f:\mathbb{R}^{n}\times\mathcal{U}\times\mathcal{D}\rightarrow\mathbb{R}^{n}$ is uniformly continuous, bounded, and Lipschitz continuous in $z$ for fixed $u$ and $d$, and $u\left(\cdot\right)\in\mathcal{U}$, $d\left(\cdot\right)\in\mathcal{D}$ are measurable functions. Under these assumptions we can guarantee the dynamical system \eqref{eq:syst-gen} has a unique solution.

In this differential game, the goal of player 2 (the disturbance) is to drive the system into some target set using only non-anticipative strategies, while player 1 (the control) aims to drive the system away from it.

We introduce the \textit{time-to-reach} problem as follows.

\textit{\textbf{(Time-to-reach) } Find the time to reach a target $\Gamma$ from any initial state $x$, in a scenario where  player 1 maximizes the time, while player 2 minimizes the time.
Player 2 is restricted to using \textit{non-anticipative} strategies,  with knowledge of player 1's current and past decisions. 
Such a time is denoted by $\phi\left(x\right)$.}

Following \cite{Yang2013}, the time to reach a closed target $\Gamma \subset \RR$ with compact boundary, given $u\left(\cdot\right)$ and $d\left(\cdot\right)$ is defined as
\[
T_{x}\left[u,d\right]=\min\left\{ t|\; z\left(t\right)\in\Gamma\right\},
\]
and the set of non-anticipative strategies as
\begin{equation*}
\begin{aligned}
\Theta=\{\theta:\mathcal{U} & \rightarrow\mathcal{D}|\; u\left(\tau\right)=\hat{u}\left(\tau\right)\,\forall\tau\le t\,\Rightarrow \\
&\,\theta\left[u\right]\left(\tau\right)=\theta\left[\hat{u}\right]\left(\tau\right)\,\forall\tau\le t \}.
\end{aligned}
\end{equation*}

Given the above definitions, the \textit{Time-to-reach} problem is equivalent to the differential game problem as follows:
\[
\phi\left(x\right)=\underset{\theta\in\Theta}{\min}\,\underset{u\in\mathcal{U}}{\max}\,T_{x}\left[u,\theta\left[u\right]\right].
\]

The collection of all the states that are reachable in a finite time is the capturability set $\mathcal{R}^{*}=\left\{ x\in\mathbb{R}^{n}|\; \phi\left(x\right)<+\infty\right\} $.

Applying the dynamic programming principle, we can obtain $\phi$ as the viscosity solution for the following stationary HJ PDE:
\begin{align}\label{eq:HJPDE}
\underset{u\in\mathcal{U}}{\min}\,\underset{d\in\mathcal{D}}{\max}\,\left\{ -\nabla\phi\left(z\right)\cdot f\left(z,u,d\right)-1\right\}  & =0,\quad\text{in}\,\mathcal{R}^{*}\backslash\Gamma\\
\phi\left(x\right) & =0,\quad\text{on\,}\Gamma. \nonumber
\end{align}

Previously, this PDE has typically been solved using  finite difference methods such as the Lax-Friedrichs method (\cite{Yang2013}).

From the solution $\phi\left(x\right)$ we can obtain the optimal avoidance control as
\begin{align}\label{eq:optimalControl}
u^{*}\left(x\right)=\arg\underset{u\in\mathcal{U}}{\min}\,\underset{d\in\mathcal{D}}{\max}\left\{-\nabla\phi\left(z\right)\cdot f\left(z,u,d\right)-1\right\}.
\end{align}


\section{Problem Formulation}
\label{sect:problem}
We consider a group of $N$ vehicles, each of them denoted as $\Qi$, $i=1,\dots,N$, with dynamics described by
\begin{equation} \label{eq:problemDynamic}
\begin{split}
\dot{p}_{i} & =v_{i},\\
\dot{v}_{i} & =u_{i},\\
\left\Vert v_{i} \right\Vert & \leq v_{max},\\
\left\Vert u_{i} \right\Vert & \leq u_{max}.
\end{split}
\end{equation}
Here, $p_{i}=\left(\pix,\piy\right)$ and $v_{i}=\left(\vix,\viy\right)$ are the position and velocity of  $\Qi$ respectively, and  $u_{i}=\left(\uix,\uiy\right)$ is the control force applied to this mobile agent.

We consider a vehicle safe if there is no other vehicle closer than a predefined collision radius $c_{r}$, in other words, if
\begin{equation}
    \centering
    \left\Vert p_{i}-p_{j} \right\Vert>c_{r}, \qquad  \text{ for any } j \neq i.
    \label{eq:safety}
\end{equation}
\vspace{0.1em}
\begin{defn} 
\label{defn:subcover}
\textbf{\textit{(r-Subcover).}} A group of agents is an \textbf{$r$-subcover} for a compact domain $\Omega\subseteq\RR$ if:
\begin{enumerate}
\item The distance between any two vehicles is at least $r$.
\item The signed distance from any vehicle to $\Omega$ is less than equal to $-\frac{r}{2}$.
\end{enumerate}
\end{defn}

\begin{defn} 
\label{defn:cover}
\textbf{\textit{(r-Cover).}} An $r$-subcover for $\Omega$ is an \textbf{$r$-cover} for $\Omega$ if its size is maximal (i.e., no larger number of agents can be an $r$-subcover for $\Omega$). 
\end{defn}

The $r$-subcover definition is closely related to finding a way to pack circular objects of radius $\frac{r}{2}$ inside of a container with shape $\Omega$. Having an $r$-cover implies the container is full and there is no room for more of such objects.

We are interested in the following safe domain coverage problem.

\textbf{\textit{(Safe-domain-coverage)}} \textit{Consider a compact domain $\Omega$ in the plane and $N$ vehicles each with dynamics described by \eqref{eq:problemDynamic}, starting from safe initial conditions. Find the maximal $r>0$ and a control policy that leads to a stable steady state which is an $r$-cover for $\Omega$, while satisfying the safety condition \eqref{eq:safety} at any time.}

\section{Methodology}
The controller we design and present has two components. First, each vehicle in the group evolves according to a coverage controller that consists in interaction forces with the rest of the vehicles and with the boundary of the target domain, as well as in braking forces in the current direction of motion. Second, a safety controller, based on Hamilton-Jacobi reachability theory, activates when two vehicles come within an unsafe region with respect to each other. The desired distance $r$ is built into the coverage controller by means of artificial potentials. For certain setups we prove that our proposed coverage control strategy is asymptotically stable, which leads to an $r$-cover for the domain when this is admissible.

\subsection{Coverage Controller}
\label{subsect:coverage-controller}
\begin{figure}
\centering
    \begin{subfigure}[b]{0.8\linewidth}        \includegraphics[width=\linewidth]{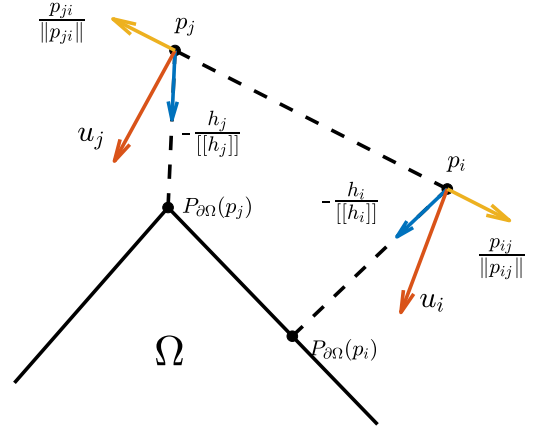}\hspace{-0.1cm}
    \end{subfigure}
\caption{Illustration of control forces acting on two vehicles located at $p_i$ and $p_j$.} 
\label{fig:controlForce}
\end{figure}

For notational simplicity, let $p_{ij}:=p_{i}-p_{j}$, denote by $h_{i}$ the shortest vector connecting $p_{i}$ with $\partial\Omega$, i.e. $h_{i}=p_{i}-P_{\partial\Omega}\left(p_{i}\right)$, and let $\left[\left[ h_{i}\right] \right] :=b\left(p_{i}\right)$. The proposed control force is given as
\begin{align}\label{eq:controlExplicit}
u_{i}=&-\sum_{j\neq i}^{N}f_{I}\left(\left\Vert p_{ij}\right\Vert \right)\frac{p_{ij}}{\left\Vert p_{ij}\right\Vert }-f_{h}\left(\left[\left[h_{i}\right] \right]\right)\frac{h_{i}}{\left[\left[h_{i}\right] \right] }\\
&+f_{v_i}\left(\left\Vert v_{i}\right\Vert \right)\frac{v_{i}}{\left\Vert v_{i}\right\Vert}, \nonumber
\end{align}
where the three terms in \eqref{eq:controlExplicit} represent inter-vehicle, vehicle-domain, and braking forces, respectively. 
Figure \ref{fig:controlForce} illustrates the control forces for two generic vehicles located at $p_i$ and $p_j$. Shown there are the unit vectors in the directions of the inter-vehicle and vehicle-domain forces (yellow and blue arrows, respectively), along with the resultant that gives the overall control force (red arrows). Note that due to the nonsmoothness of the boundary, different points may have different types of projections: $p_i$ projects on the foot of the perpendicular to $\partial \Omega$, while $p_j$ projects on a corner point of $\partial \Omega$.

\begin{figure}
\centering
    \begin{subfigure}[b]{4.2cm}
        \includegraphics[width=4.2cm]{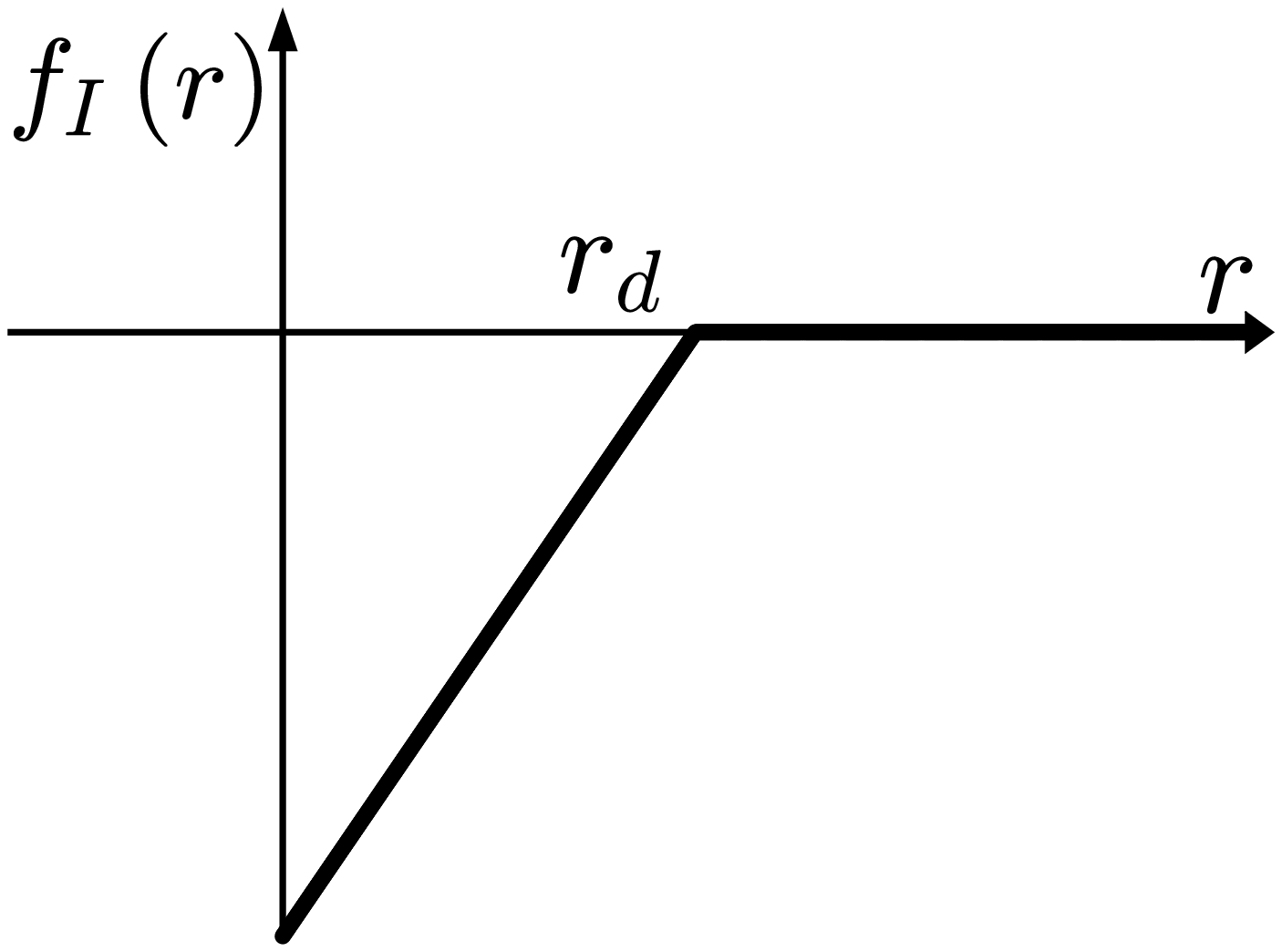}\hspace{-0.1cm}
    \end{subfigure}
    \begin{subfigure}[b]{4.2cm}
        \includegraphics[width=4.2cm]{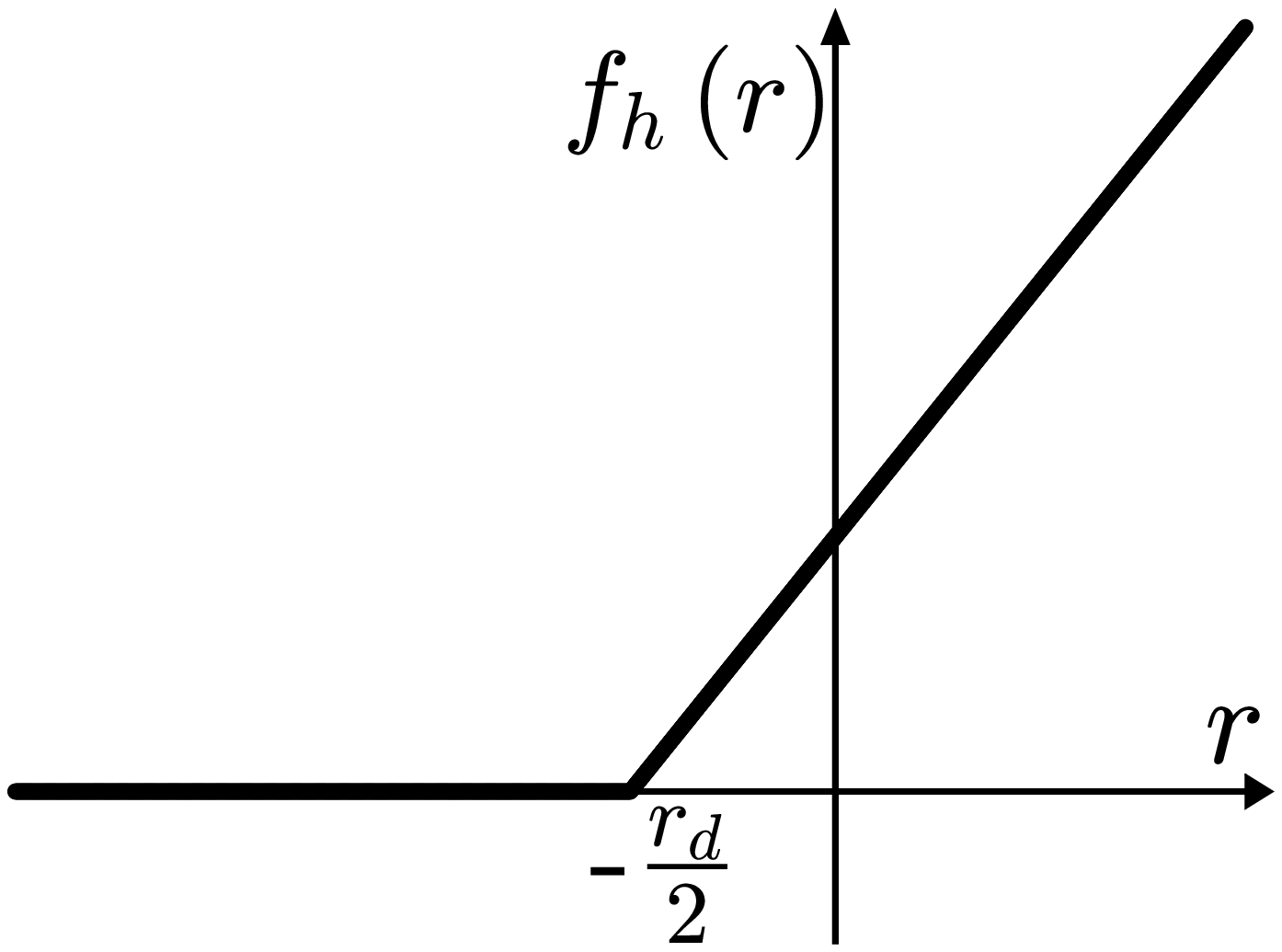}
    \end{subfigure}
\caption{Inter-vehicle and vehicle-domain control forces.} 
\label{fig:fINfh}
\end{figure}

We assume the following forms for the functions $f_{I}, f_{h}$ and $f_{v_i}$ that appear in the various control forces in Equation \eqref{eq:controlExplicit}. Figure \ref{fig:fINfh} shows the inter-vehicle force $f_{I}$ and the vehicle-domain force $f_{h}$. Note that $f_{I}(r)$ is negative for $r<\rd$, and zero otherwise. This means that for two vehicles within distance $0<r<\rd$ from each other, their inter-vehicle interactions are repulsive, while two vehicles at distance larger than $\rd$ apart do not interact at all. The vehicle-domain force $f_{h}(r)$ is zero for $r<-\frac{\rd}{2}$, and positive for $r>-\frac{\rd}{2}$.  For a vehicle $i$ outside the target domain, i.e., with $\left[\left[h_{i}\right] \right] >0$, this results in an attractive interaction force toward $\partial \Omega$. On the other hand, for a vehicle inside the domain, where $\left[\left[h_{i}\right] \right] <0$, one distinguishes two cases: i) the vehicle is within distance $\frac{\rd}{2}$ to the boundary, in which case it experiences a repulsive force from it, or ii) the vehicle is more than distance $\frac{\rd}{2}$ from the boundary, in which case it does not interact with the boundary at all. Finally, we take $f_{v_i}(\|v_i\|)$ to be negative  to result in a braking force; a specific form of $f_{v_i}$ will be chosen below for analytical considerations (see Equation \eqref{eq:fv}).

An important ingredient of our controller is that one can associate a Lyapunov function to it and hence, investigate analytically the stability of its solutions. We address these considerations now.

\begin{lem}
\label{lemma:potential}
The vehicle-domain force $-f_{h}\left(\left[\left[h_{i}\right] \right] \right)\frac{h_{i}}{\left[\left[h_{i}\right] \right] }$ and the inter-vehicle  force $-f_{I}\left(\left\Vert p_{ij}\right\Vert \right)\frac{p_{ij}}{\left\Vert p_{ij}\right\Vert }$  are conservative.
\end{lem}

\begin{pf}
 Let us consider the potential 
 \begin{align*}
     V_{h}\left(p_{i}\right)={\int_{-\frac{r_{d}}{2}}^{\left[\left[ h_{i}\right]\right] }} f_{h} \left(s\right) ds,
 \end{align*}
which satisfies
\begin{align*}
\nabla_{i} V_{h}\left(p_{i}\right)&=f_{h}\left(\left[\left[h_{i}\right] \right] \right)\nabla(\left[\left[h_{i}\right] \right])\\
&=f_{h}\left(\left[\left[h_{i}\right] \right] \right)\frac{h_{i}}{\left[\left[h_{i}\right] \right] }.
\end{align*}
where we have used Equations \eqref{eq:sgnDistChar} and \eqref{eq:sgnDistGrad} for $\left[\left[h_{i}\right] \right]$ and $\nabla(\left[\left[h_{i}\right] \right])$ respectively (see Theorem 5.1(iii) in \cite{DelfourZolesio2001}).

Similarly, it can be shown that the inter-vehicle force is the negative gradient of the potential
\begin{align*}
V_{I}\left(p_{ij}\right)={\int_{r_{d}}^{\left\Vert p_{ij}\right\Vert }}f_{I}\left(s\right)ds.
\end{align*}
\qedwhite
\end{pf}

Using Lemma \ref{lemma:potential}, the control given in Equation \eqref{eq:controlExplicit} becomes
\begin{equation}\label{eq:controlPotential}
u_{i}={\sum_{j\neq i}^{N}}-\nabla_{i}V_{I}\left(p_{ij}\right)-\nabla_{i}V_{h}\left(p_{i}\right)+f_{v_i}\left(\left\Vert v_{i}\right\Vert \right)\frac{v_{i}}{\left\Vert v_{i}\right\Vert}.
\end{equation}

\textit{Asymptotic behaviour of the controlled system.} Consider the following candidate for a Lyapunov function, consisting in kinetic plus (artificial) potential energy:
\[
\Phi=\frac{1}{2}{\sum_{i=1}^{N}}\Bigl(\dot{p}_{i}\cdot\dot{p}_{i}+{\sum_{j\neq i}^{N}}V_{I}\left(p_{ij}\right)+2V_{h}\left(p_{i}\right)\Bigr).
\]
Note that each term in $\Phi$ is non-negative, and $\Phi$ reaches its absolute minimum value when the vehicles are totally stopped. Also, at the global minimum $\Phi=0$, the equilibrium configuration is an $\rd$-subcover of $\Omega$; in particular, all vehicles are inside the target domain.


The derivative of $\Phi$ with respect to time can be calculated as:
\begin{align*}
\dot{\Phi} & ={\sum_{i=1}^{N}}\dot{p}_{i}\cdot\Bigl(u_{i}+{\sum_{j\neq i}^{N}}\nabla_{i}V_{I}\left(p_{ij}\right)+\nabla_{i}V_{h}\left(p_{i}\right)\Bigr)\\
 & ={\sum_{i=1}^{N}}\dot{p}_{i}\cdot f_{v_i}\left(\left\Vert v_{i}\right\Vert \right)\frac{v_{i}}{\left\Vert v_{i}\right\Vert }\\
 & ={\sum_{i=1}^{N}}f_{v_i}\left(\left\Vert v_{i}\right\Vert \right)\left\Vert v_{i}\right\Vert,
\end{align*}
where we used the dynamics \eqref{eq:problemDynamic} and equation \eqref{eq:controlPotential}. Thus, if we choose
\begin{equation}
\label{eq:fv}
f_{v_i}(\left\Vert v_{i}\right\Vert)=-a_i \left\Vert v_{i}\right\Vert, \quad \text { with } a_i>0, \quad i=1,\dots,N,
\end{equation}
then $\dot{\Phi}$ is negative semidefinite and equal to zero if and only if $\dot{p_{i}}=0$ for all $i$ (i.e., all vehicles are at equilibrium). By LaSalle Invariance Principle we can conclude that the controlled system approaches asymptotically an equilibrium configuration.

For certain simple setups (e.g., a square number of vehicles in a square domain or a triangular number of vehicles in a triangular domain -- see Figs. \ref{fig:squareSteady} and \ref{fig:triangleSteady}), the $r_{d}$-covers are isolated equilibria. Hence, together with the fact that such equilibria are global minimizers for $\Phi$, their local asymptotic stability can be inferred. The formal result is given by the following proposition.

\begin{prop}
\label{prop:stability}
Consider a group of $N$ vehicles with dynamics defined by \eqref{eq:problemDynamic}, and the control law given by \eqref{eq:controlPotential} and \eqref{eq:fv}. Let the equilibrium of interest be of the form $\dot{p_{i}}=0$,  $\left\Vert p_{ij}\right\Vert \geq r_{d}$  and $\left[\left[h_{i}\right]\right]\leq -\frac{r_{d}}{2}$ for  $i,j=1,\cdots,N$ (see Definitions \ref{defn:subcover} and \ref{defn:cover}), and assume that this equilibrium configuration is isolated. Also assume that there is a neighborhood about the equilibrium in which the control law remains smooth. Then, the equilibrium is a global minimum of the sum of all the artificial potentials and is locally asymptotically stable.
\end{prop}

Choosing an adequate $r_{d}$ when solving the \textit{safe-domain-coverage} problem leads us to a nonlinear optimization problem (see \cite{lopez2011heuristic}) which can be difficult in itself. Our heuristic approach for picking this parameter relies on the premise that any vehicle is covering roughly the same square area, i.e.,
\begin{equation}
\label{eqn:rd}
    r_d = \sqrt{\frac{Area\left(\Omega\right)}{N}}.
\end{equation}
Note that \eqref{eqn:rd} gives the exact formula for the maximal radius when both the number of vehicles and the domain are square, making  the $r_{d}$-covers isolated equilibria, a key assumption in Proposition \ref{prop:stability}. Also, this particular choice for $r_d$ leads to the desired cover for several domains with regular geometries (see Section \ref{sect:numerics}).


\subsection{Collision Avoidance via Analytic HJ PDE Solution}
\label{subsect:collision-controller}

The dynamics between two vehicles $\Qi$, $\Qj$ can be defined in terms of their relative states
\begin{align*}
\pxr & =\pix-\pjx\\
\pyr & =\piy-\pjy\\
\vxr & =\vix-\vjx\\
\vyr & =\viy-\vjy,
\end{align*}

\noindent where the vehicle $\Qi$ is the evader and $\Qj$ is the pursuer, the latter being considered as the model disturbance. In this case the relative dynamical system is given by
\begin{align}
\label{eq:relativeDynSys}
\begin{split}
\pdxr & = \vxr\\
\pdyr & = \vyr\\ 
\vdxr & = \uix-\ujx\\
\vdyr & = \uiy-\ujy\\
\left\Vert u_{i}\right\Vert,\left\Vert u_{j}\right\Vert & \leq u_{max},
\end{split}
\end{align}

\noindent where $u_i=\left(\uix,\uiy\right)$ and $u_j=\left(\ujx,\ujy\right)$ are the control inputs of the agents $\Qi$ and $\Qj$, respectively. 
From the perspective of agent $\Qi$, the control inputs of $\Qj$, $u_j=\left(\ujx,\ujy\right)$, are treated as worst-case disturbance.

According to \eqref{eq:safety}, the unsafe states are described by the target set
\begin{align*}
\Gamma=\left\{ z:\pxr^{2}+\pyr^2\leq c_{r}^{2}\right\}.
\end{align*}

Consider the time it takes for the solution of the dynamical system \eqref{eq:relativeDynSys}, with starting point $z$ in $\mathcal{R^{*}}\setminus\Gamma$, to reach $\Gamma$ when the disturbance and control inputs are optimal. 
Using a geometric argument one can show that this time is the minimum of the two solutions of the following quadratic equation:
\begin{align}\label{eq:quadraticTTR}
\left(\vxr^{2}+\vyr^{2}\right)\ttr^{2}\left(z\right)&+2\left(\pxr\vxr+\pyr\vyr\right)\ttr\left(z\right)\\
&+\left(\pxr^{2}+\pyr^{2}-c_{r}^{2}\right)=0.\nonumber
\end{align}

Let us verify first that the minimum of the two solutions satisfies indeed the HJ PDE \eqref{eq:HJPDE}.
\begin{prop}
Consider the function $\ttr\left(z\right)$ defined as
\[
\ttr\left(z\right) := \frac{-\left(\pxr\vxr+\pyr\vyr\right) - \sqrt{\Delta}}{\vxr^{2}+\vyr^{2}} \qquad \text{ in } \mathcal{R^{*}}\setminus\Gamma,
\]
where 
\[
\Delta = \left(\pxr\vxr+\pyr\vyr\right)^{2}-\left(\vxr^{2}+\vyr^{2}\right)\left(\pxr^{2}+\pyr^{2}-c_{r}^{2}\right).
\]
Also define $\ttr\left(z\right)$ to be $0$ on $\Gamma$. Then $\ttr\left(z\right)$ satisfies equation \eqref{eq:HJPDE}.
\end{prop}

\begin{pf} For $z\in\mathcal{R^{*}}\setminus\Gamma$, $\ttr\left(z\right)$ is the minimum solution of the quadratic equation \eqref{eq:quadraticTTR}. By using implicit differentiation on \eqref{eq:quadraticTTR} one can show
\begin{align*}
\frac{\partial\ttr}{\partial \pxr} =-\frac{\vxr\ttr\left(z\right)+\pxr}{\left(\vxr^{2}+\vyr^{2}\right)\ttr\left(z\right)+\left(\pxr\vxr+\pyr\vyr\right)}\\
\frac{\partial\ttr}{\partial \pyr} =-\frac{\vyr\ttr\left(z\right)+\pyr}{\left(\vxr^{2}+\vyr^{2}\right)\ttr\left(z\right)+\left(\pxr\vxr+\pyr\vyr\right)}.
\end{align*}

To put system \eqref{eq:relativeDynSys} in the general form \eqref{eq:syst-gen} from Sec. \ref{subsect:HJ-reach}, let $z=(\pxr,\pyr,\vxr,\vyr)$, $u=\left(u_x,u_y\right):=\left(\uix,\uiy\right)$, $d=\left(d_x,d_y\right):=\left(\ujx,\ujy\right)$, and let $f\left(z,u,d\right)$ represent the right-hand-side of \eqref{eq:relativeDynSys}. Then,
\begin{align*}
&\underset{u\in\mathcal{U}}{\min}\,\underset{d\in\mathcal{D}}{\max}\left\{ -\nabla\ttr\left(z\right)\cdot f\left(z,u,d\right)-1\right\}  \\
&=\underset{u\in\mathcal{U}}{\min}\,\underset{d\in\mathcal{D}}{\max}\left\{ -\frac{\partial\ttr\left(z\right)}{\partial \pxr}\vxr-\frac{\partial\ttr\left(z\right)}{\partial \vxr}\left(u_{x}-d_{x}\right)\right. \\
&\qquad\qquad\qquad\left.-\frac{\partial\ttr\left(z\right)}{\partial \pyr}\vyr-\frac{\partial\ttr\left(z\right)}{\partial \vyr}\left(u_{y}-d_{y}\right)-1\right\} \\
&=-\left(\frac{\partial\ttr\left(z\right)}{\partial \pxr}\vxr+\frac{\partial\ttr\left(z\right)}{\partial \pyr}\vyr\right)-1\\
&=\frac{\left(\vxr\ttr\left(z\right)+\pxr\right)\vxr+\left(\vyr\ttr\left(z\right)+\pyr\right)\vyr}{\left(\vxr^{2}+\vyr^{2}\right)\ttr\left(z\right)+\left(\pxr\vxr+\pyr\vyr\right)}-1\\
&=0,
\end{align*}
where we have used the fact that $\ttr\left(z\right)$ is differentiable in $\mathcal{R^{*}}\setminus\Gamma$ and that the minimum and maximum in the second equal sign are attained at
\begin{subequations}
\begin{equation}\label{eq:optimalControlForm1}
    u^{*}=u_{max}\frac{\left(\frac{\partial\ttr\left(z\right)}{\partial \vxr},\frac{\partial\ttr\left(z\right)}{\partial \vyr}\right)}
    {\left\Vert\frac{\partial\ttr\left(z\right)}{\partial \vxr},\frac{\partial\ttr\left(z\right)}{\partial \vyr}\right\Vert} 
\end{equation}
\begin{equation}
    d^{*}=u_{max}\frac{\left(\frac{\partial\ttr\left(z\right)}{\partial \vxr},\frac{\partial\ttr\left(z\right)}{\partial \vyr}\right)}
    {\left\Vert\frac{\partial\ttr\left(z\right)}{\partial \vxr},\frac{\partial\ttr\left(z\right)}{\partial \vyr}\right\Vert}
\end{equation}
\end{subequations}

\noindent and therefore cancel out.

For $z\in\Gamma$ the equation \eqref{eq:HJPDE} is satisfied by the definition of $\ttr$.

\qedwhite
\end{pf}

Similarly as before, by implicit differentiation of \eqref{eq:quadraticTTR} one can also show
\begin{subequations}
\begin{align}
    \frac{\partial\ttr}{\partial \vxr} =-\frac{\vxr\ttr^{2}\left(z\right)+\pxr\ttr\left(z\right) }{\left(\vxr^{2}+\vyr^{2}\right)\ttr\left(z\right)+\left(\pxr\vxr+\pyr\vyr\right)}\label{eq:dpsi_dvx}
    \\
    \frac{\partial\ttr}{\partial \vyr} =-\frac{\vyr\ttr^{2}\left(z\right)+\pyr\ttr\left(z\right) }{\left(\vxr^{2}+\vyr^{2}\right)\ttr\left(z\right)+\left(\pxr\vxr+\pyr\vyr\right)}.\label{eq:dpsi_dvy}
\end{align}

\end{subequations}

Now, by using \eqref{eq:optimalControlForm1} we can derive a closed expression for the optimal avoidance controller.

For applications using the static HJ PDE \eqref{eq:HJPDE} or similar equations, the solution is commonly approximated via finite difference methods such as the one presented in \cite{Yang2013}; however, using an analytic solution leads to two main advantages. 
First, refinements in the resolution when using uniform grid point spacing may require a large amount of memory and long computational times that scale very poorly. Second, these methods are only able to compute the solution in a bounded domain. Having an analytical solution allows us to have the best possible resolution in an unbounded domain. Practically, this allows us to predict and react to possible collisions arbitrarily far into the future.


\subsection{Overall Control Logic}
\label{subsect:logic}
We have presented two controllers. First, a controller based on virtual potentials which leads to coverage, but without taking into consideration safety; and second, a safety controller that guarantees pairwise collision avoidance. The main objective of this subsection is to describe how to switch between these two controllers.

We will consider that vehicle $\Qi$ is  in potential conflict with vehicle $\Qj$ if the time to collision $\ttr\left(z_{i}\right)$ (time to reach $\Gamma$), given the relative current state $z_{i}$, is less than or equal to a specified time horizon $t_{safety}$. In other words, a danger is defined when a vehicle is susceptible to collision in the next $t_{safety}$ seconds. In such a case $\Qi$ must use the safety controller, otherwise, the coverage controller is used.

In the case that a vehicle detects more than one conflict, it will apply the control policy of the first conflict detected at that particular time. 
Algorithm \ref{alg:overallControlLogic} describes the overall control logic for every vehicle $\Qi$.

\IncMargin{2 em}
\begin{algorithm2e}[ht]
\SetAlgoLined
\KwIn{State $x_i$ of a vehicle $\Qi$; states $\{x_j\}_{j\neq i}$ of other vehicles $\left\{ Q_j\right\}_{j\neq i}$; a domain $\Omega$ to cover.}
\nonl\textbf{Parameter:} A time horizon for safety check $t_{safety}$\;
\KwOut{A control $u_i$ for $\Qi$.}
 \vspace{1em}
 $safe \leftarrow \text{True}$\;
 \For{$j\neq i$}{
    $z \leftarrow x_{i}-x_{j}$\;
    \If{$\ttr\left(z\right) \leq t_{safety}$}{
        $safe \leftarrow \text{False}$\;
        $U_{ix}=-\frac{\vxr\ttr^{2}\left(z\right)+\pxr\ttr\left(z\right)}{\left(\vxr^{2}+\vyr^{2}\right)\ttr\left(z\right)+\left(\pxr\vxr+\pyr\vyr\right)}$\;\label{alg:ln:avoidUx}
        $U_{iy}=-\frac{\vyr\ttr^{2}\left(z\right)+\pyr\ttr\left(z\right)}{\left(\vxr^{2}+\vyr^{2}\right)\ttr\left(z\right)+\left(\pxr\vxr+\pyr\vyr\right)}$\;\label{alg:ln:avoidUy}
        \textbf{break for}
    }
 }
  
  \If{safe}{
        $\left(U_{ix},U_{iy}\right)=\text{-}\sum_{j\neq i}^{N}f_{I}\left(\left\Vert p_{ij}\right\Vert \right)\frac{p_{ij}}{\left\Vert p_{ij}\right\Vert }\text{-}f_{h}\left(\left[\left[h_{i}\right] \right]\right)\frac{h_{i}}{\left[\left[h_{i}\right] \right]}+f_{v_i}\left(\left\Vert v_{i}\right\Vert \right)\frac{v_{i}}{\left\Vert v_{i}\right\Vert}$\;\label{alg:ln:coverU}
   }{
  }
  $u_{i}=u_{max}\frac{\left(U_{ix},U_{iy}\right)}{\left\Vert\left(U_{ix},U_{iy}\right)\right\Vert}$\;\label{alg:ln:normalizing}
  \KwRet{$u_{i}$};
 \caption{Overall control logic for a generic vehicle $\Qi$. }
 \label{alg:overallControlLogic}
\end{algorithm2e}
\DecMargin{0em}

In Algorithm \ref{alg:overallControlLogic}, lines \ref{alg:ln:avoidUx} and \ref{alg:ln:avoidUy} can be obtained from equations \eqref{eq:dpsi_dvx}, \eqref{eq:dpsi_dvy} and \eqref{eq:optimalControlForm1} (also note the normalization step in line \ref{alg:ln:normalizing}), while line \ref{alg:ln:coverU} comes from the explicit coverage control \eqref{eq:controlExplicit}. 


\section{Numerical Simulations}
\label{sect:numerics}
In this section, we show numerical simulations with various domains  using the coverage and safety controllers discussed above.

\subsection{Covering a Square Domain}
\label{subsect:square}
We first consider the problem in which several vehicles cover a square domain. Here we present two strategies: while both of them use the coverage controller described in Sec. \ref{subsect:coverage-controller}, only the second strategy switches to the safety controller when necessary, according to Sec. \ref{subsect:logic}. In both cases 16 vehicles start from a horizontal line setup outside of the target square domain, as shown in  Fig. \ref{fig:squareInitial}.

\begin{figure}
    \centering
    \begin{subfigure}[b]{4.2cm}
    \centering
    \includegraphics[width=4.2cm]{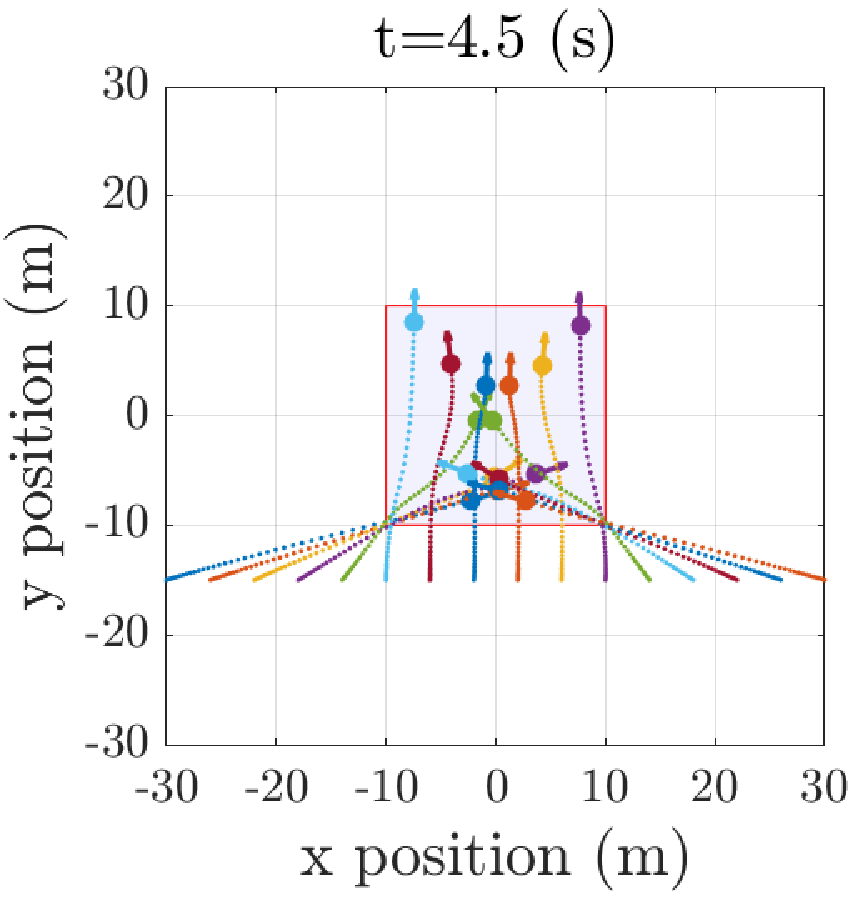}
    \caption{}
    \label{fig:squareNoAvoid4_5}
    \end{subfigure}
    \hspace{-0.2cm}
    \begin{subfigure}[b]{4.2cm}
    \centering
    \includegraphics[width=4.2cm]{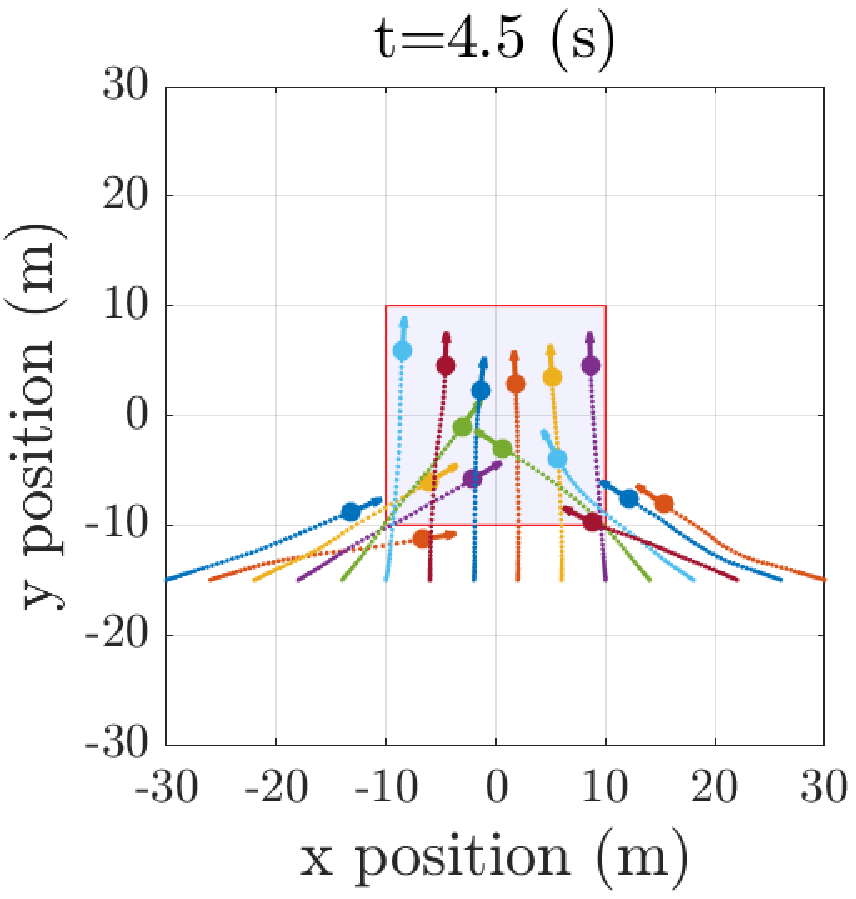}
    \caption{}
    \label{fig:squareAvoid4_5}
    \end{subfigure}
    \begin{subfigure}[b]{4.2cm}
    \includegraphics[width=4.2cm]{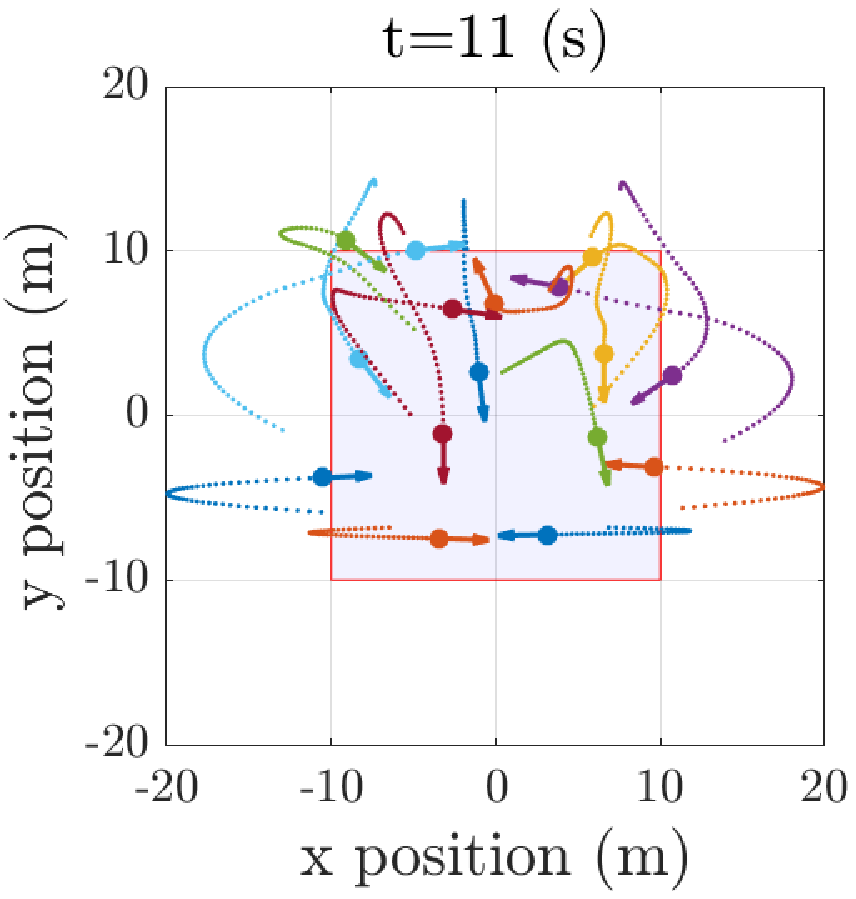}\hspace{-0.2cm}
    \caption{}
    \label{fig:squareNoAvoid11}
    \end{subfigure}
    \begin{subfigure}[b]{4.2cm}
    \includegraphics[width=4.2cm]{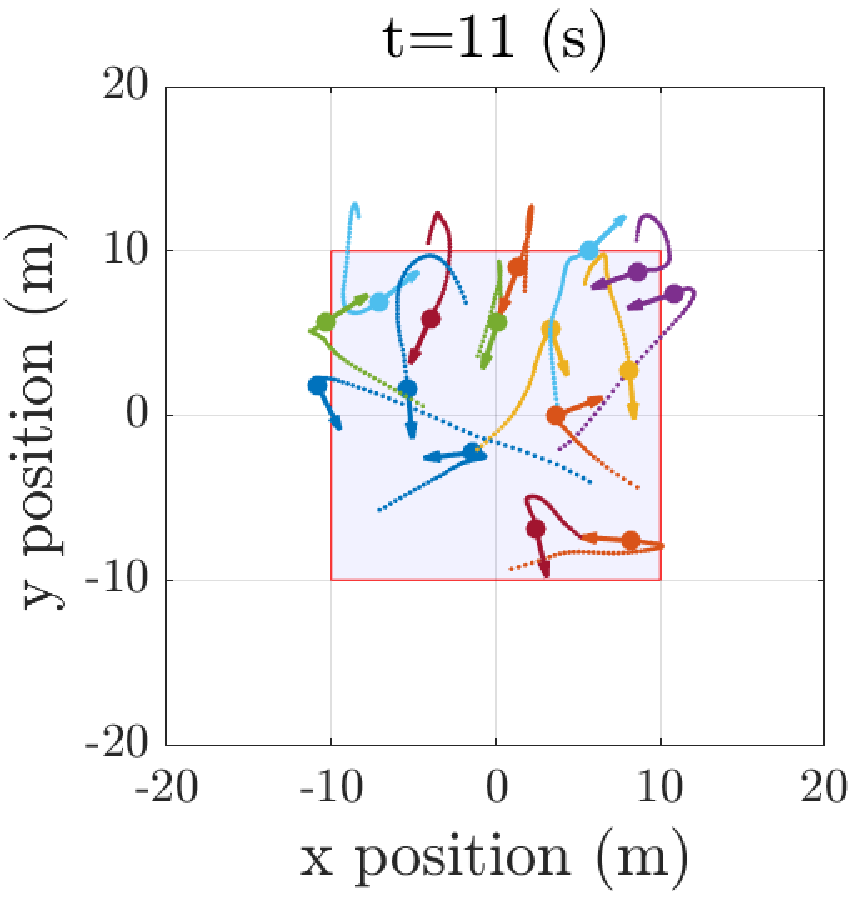}
    \caption{}
    \label{fig:squareAvoid11}
    \end{subfigure}
    \begin{subfigure}[b]{4.2cm}
    \includegraphics[width=4.2cm]{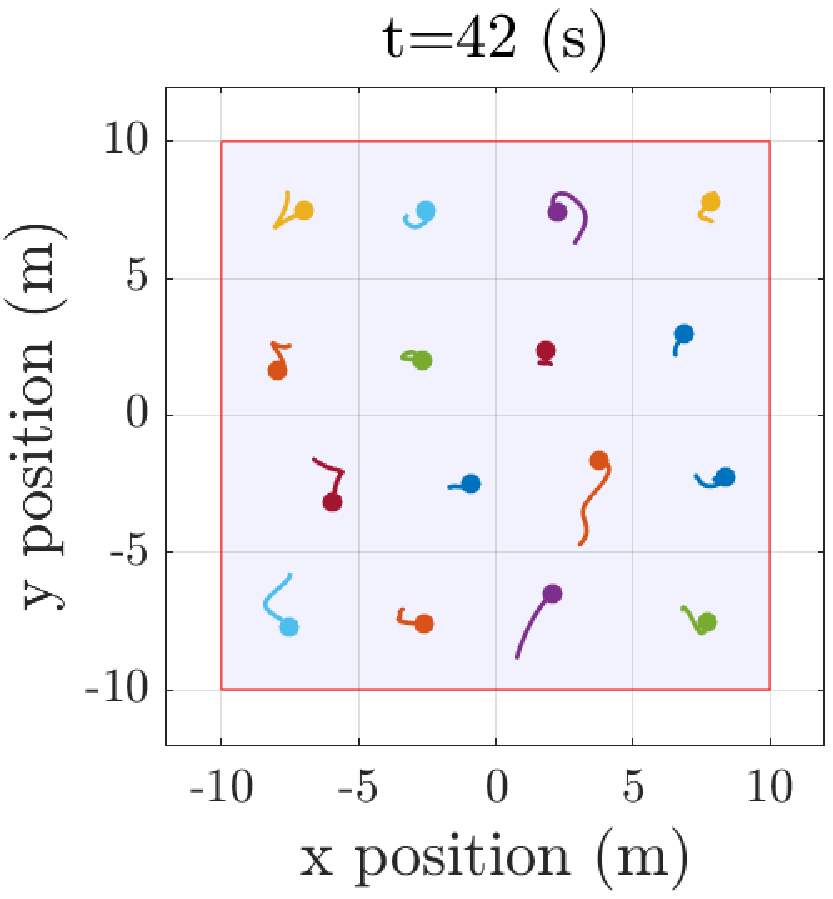}\hspace{-0.2cm}
    \caption{}
    \label{fig:squareNoAvoid43}
    \end{subfigure}
    \begin{subfigure}[b]{4.2cm}
    \includegraphics[width=4.2cm]{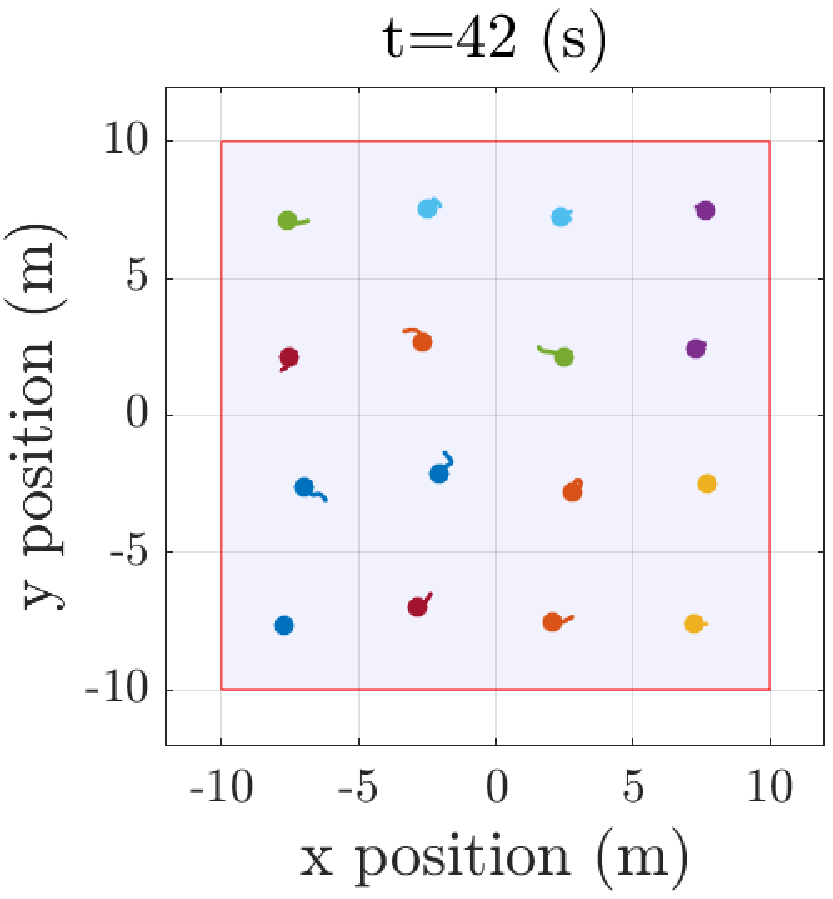}
    \caption{}
    \label{fig:squareAvoid43}
    \end{subfigure}
    \caption{Square domain coverage at different time instants, without (left) and with (right) safety controller, when $N=16$, $c_{r}=2\,\text{(m)}$, $v_{max}=10\,\text{(m/s)}$, $u_{max}=3 \,\text{(m/s$^{2})$}$, $t_{safety}=5\,\text{(s)}$, side length $l=20\,\text{(m)}$, domain area $A=l^{2}=400\,\text{(m$^2$)}$ and $r_{d}=\sqrt{\frac{A}{N}}=5\,\text{(m)}$. Vehicles start in a horizontal line configuration and reach a square grid steady state which is an $r_d$-cover of the domain (see Definition \ref{defn:cover}). The use of the safety controller reduces both the collision count and the overshoot, and helps reach the steady state faster.} 
    \label{fig:square}
\end{figure}

The left column of Fig. \ref{fig:square} illustrates different time steps for the scenario when the safety controller is not used, while the right column shows results in the presence of the safety controller. The large coloured dots represent the position of the vehicles, the dashed tails are past trajectories (shown for the previous 5 seconds), and the arrows indicate the movement direction. Note however that we have not shown the arrows when the velocities are very small, as the tails are more meaningful in this case.

At $t=0\text{ (s)}$ there are no contributions from the inter-vehicle forces or from the safety controller as the vehicles are far from each other and the initial speed is zero. The only contribution comes from the vehicle-domain forces, which pull the mobile agents toward the interior of the square; see initial trajectory tails in Figs. \ref{fig:squareNoAvoid4_5} and \ref{fig:squareAvoid4_5}. At $t=4.5\text{ (s)}$ the vehicles without safety controller are more prone to collisions due to the symmetry of the initial condition. The safety controller slows down some vehicles hence breaking this symmetry, and allowing them to enter in the crowded area without collisions.

As the forces between vehicles and domain are piece-wise linear with respect to the distance, they can be thought of as spring-like forces; in that respect, the presence of overshoots is expected, as shown in Fig. \ref{fig:squareNoAvoid11}. However, Fig. \ref{fig:squareAvoid11} indicates that in addition to preventing collisions, the use of the safety controller reduces the overshoot considerably.

After $t=60\text{ (s)}$ both control strategies reach the steady state shown in Fig. \ref{fig:squareSteady}. We note however that the system with collision avoidance reaches the equilibrium faster, as shown by Figs. \ref{fig:squareNoAvoid43} and \ref{fig:squareAvoid43}.

A \textit{collision event} starts when the distance between two vehicles is less than or equal to the collision radius $c_{r}$, and ends when the distance becomes greater than $c_{r}$. The collision event count for the square domain coverage with and without the safety controller, for various number of vehicles, is shown in Table \ref{tb:collisionsSquare}. It is noteworthy that in the absence of the safety controller the collision count increases significantly with the number of vehicles, while it remains zero when the safety controller is used.

\begin{table}[H]
    \captionsetup{width=\linewidth}
    \caption{Square coverage collision count.}
    \label{tb:collisionsSquare}
    \centering
    \begin{tabular}{ccc}
    number of vehicles & without avoidance & with avoidance\tabularnewline
    \hline
    9 & 12 & 0\tabularnewline
    16 & 39 & 0\tabularnewline
    25 & 124 & 0\tabularnewline
    \hline
    \end{tabular}
\end{table}


\subsection{Covering a Triangular Domain}
\label{subsect:triangle}

\begin{figure}
\centering
\begin{subfigure}[b]{4.2cm}
\includegraphics[width=4.2cm]{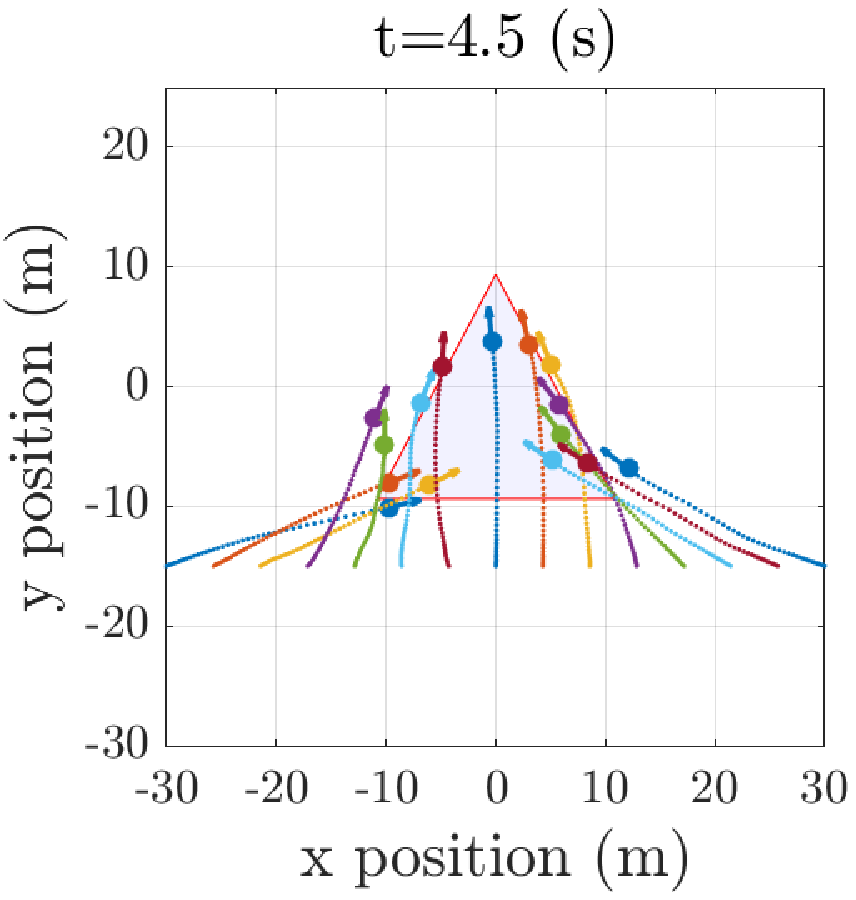}\hspace{-0.2cm}
\caption{}
\end{subfigure}
\begin{subfigure}[b]{4.2cm}
\includegraphics[width=4.2cm]{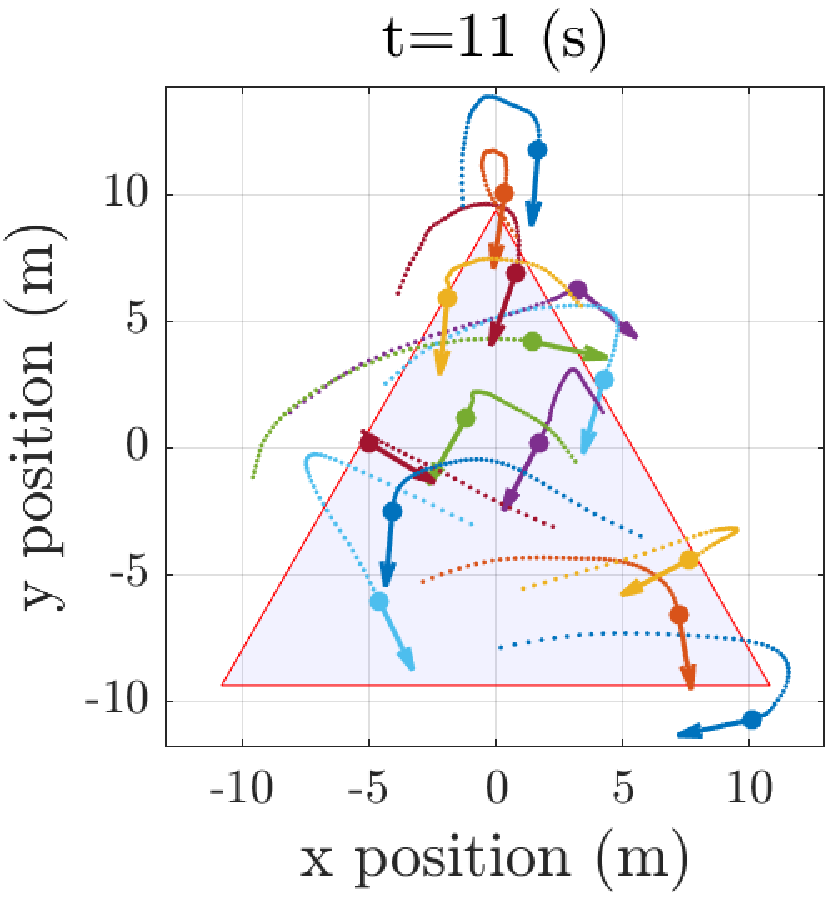}
\caption{}
\end{subfigure}
\begin{subfigure}[b]{4.2cm}
\includegraphics[width=4.2cm]{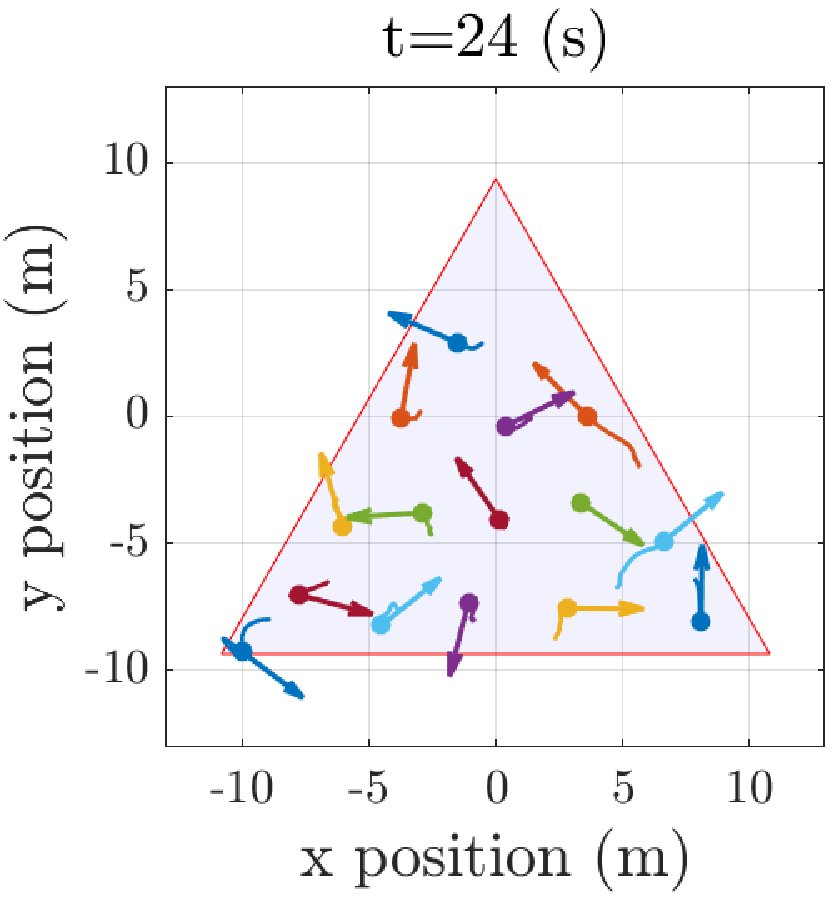}\hspace{-0.2cm}
\caption{}
\end{subfigure}
\begin{subfigure}[b]{4.2cm}
\includegraphics[width=4.2cm]{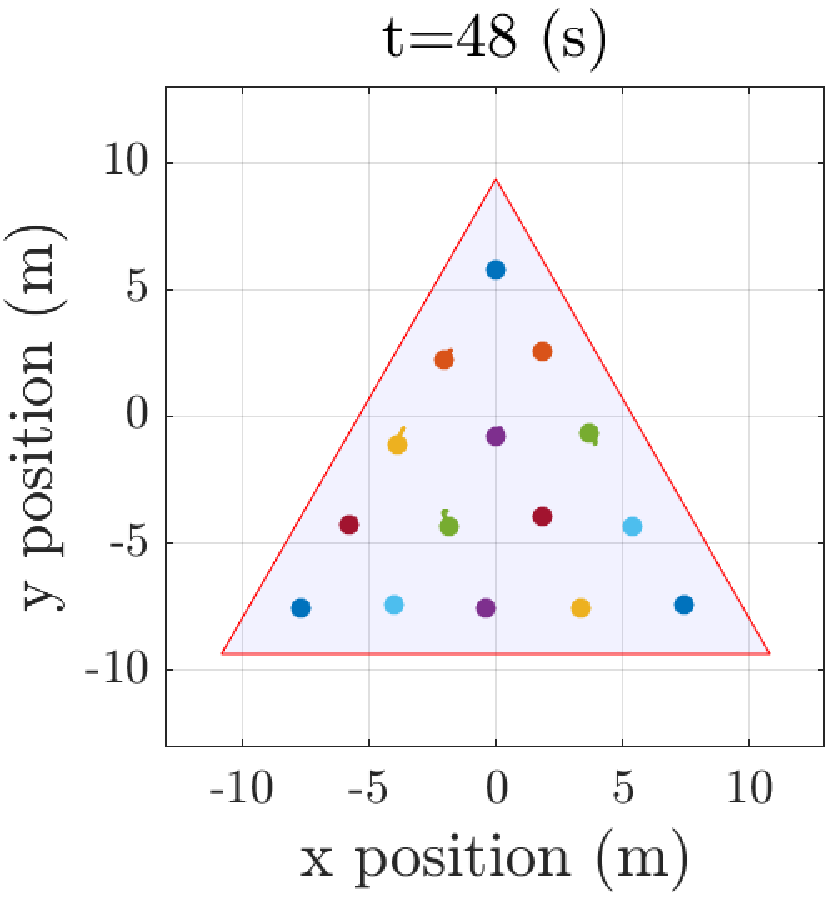}
\caption{}\label{fig:triangleSteady}
\end{subfigure}
\caption{Equilateral triangular domain coverage at different times instants, with safety controller, when $N=15$, $c_{r}=2\,\text{(m)}$, $v_{max}=10\,\text{(m/s)}$, $u_{max}=3 \,\text{(m/s$^{2})$}$, $t_{safety}=5\,\text{(s)}$, side length $l=\frac{25\sqrt{3}}{2}\text{(m)}$, domain area $A=\frac{\sqrt{3}l^{2}}{4}=\frac{1875\sqrt{3}}{16}\text{(m$^2$)}$ and $r_{d}=\sqrt{\frac{A}{N}}=\frac{5\sqrt{5\sqrt{3}}}{4}\,\text{(m)}$. The vehicles start in a horizontal line configuration and reach a steady state formation that is an $r_d$-cover of the domain.}
\label{fig:triangle}
\end{figure}

Now we consider the scenario in which an equilateral triangular domain is covered by a triangular number of vehicles, i.e. $N=\frac{n\left(n+1\right)}{2},\, n\in\mathbb{N}$. At the start of the simulation the vehicles lie on a horizontal line outside the domain. The evolution for a group of $N=15$ agents, each of them using the coverage and pairwise safety strategies discussed earlier, is illustrated in Fig. \ref{fig:triangle}. The tails represent the 5-second history of the vehicle positions.

The triangle coverage simulation was run for 6, 10 and 15 vehicles. Table \ref{tb:collisionsTriangle} shows the collision count, with and without the safety controller. Again, the use of the pairwise safety strategy reduces the number of collisions considerably. We note here that safety issues may arise when a vehicle needs to avoid more than two vehicles at the same time. In this case the pairwise safety approach used in this paper does not guarantee collision avoidance.
Guaranteed collision avoidance for more than two vehicles is explored in \cite{Chen2016}.

\begin{table}[hb]
\centering
\captionsetup{width=\linewidth}
\caption{Triangle coverage collision count}
\begin{tabular}{ccc}
number of vehicles & without avoidance & with avoidance\tabularnewline
\hline
6 & 9 & 0\tabularnewline
10 & 23 & 0\tabularnewline
15 & 72 & 2\tabularnewline
\hline
\end{tabular}
\label{tb:collisionsTriangle}
\end{table}


\subsection{Covering a Moving Non-Convex Domain}
\label{subsect:non-convex}
\begin{figure}
\centering
\begin{subfigure}[b]{4.2cm}
\includegraphics[width=4.2cm]{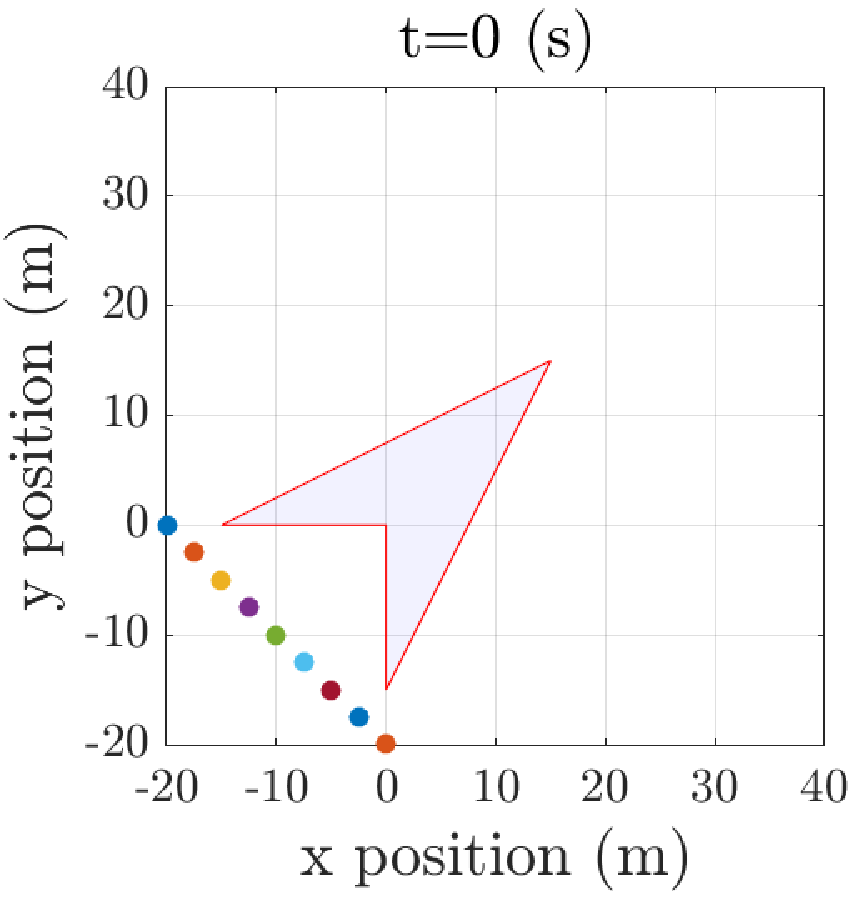}\hspace{-0.2cm}
\caption{}
\label{fig:arrow0}
\end{subfigure}
\begin{subfigure}[b]{4.2cm}
\includegraphics[width=4.2cm]{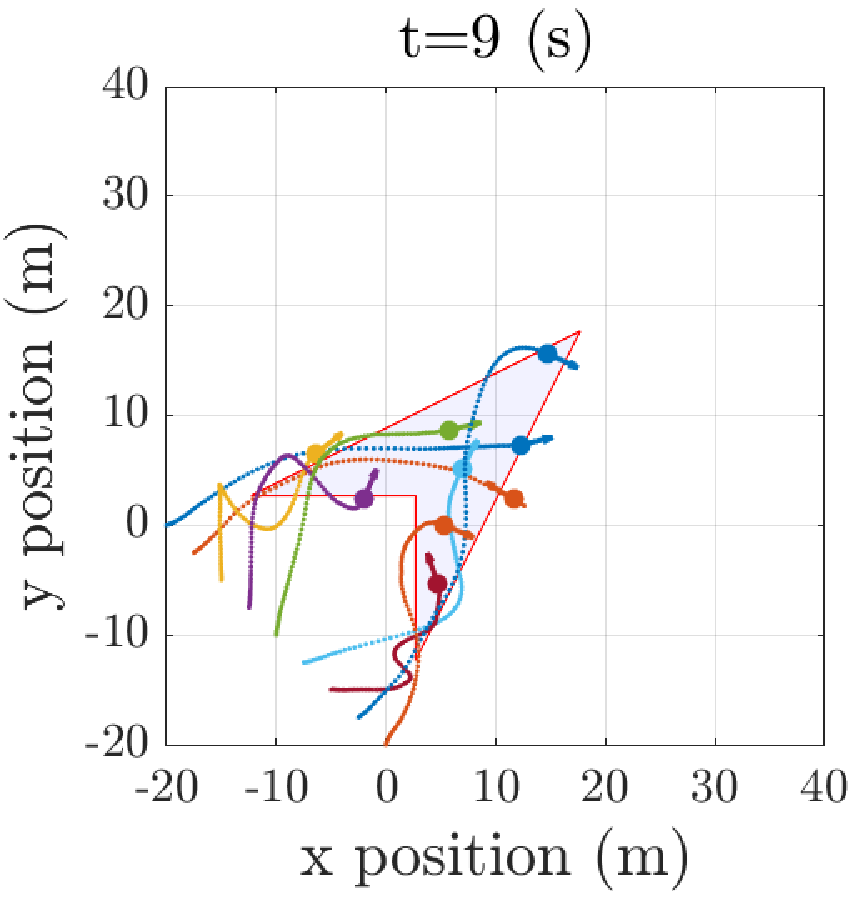}
\caption{}
\label{fig:arrow9}
\end{subfigure}
\begin{subfigure}[b]{4.2cm}
\includegraphics[width=4.2cm]{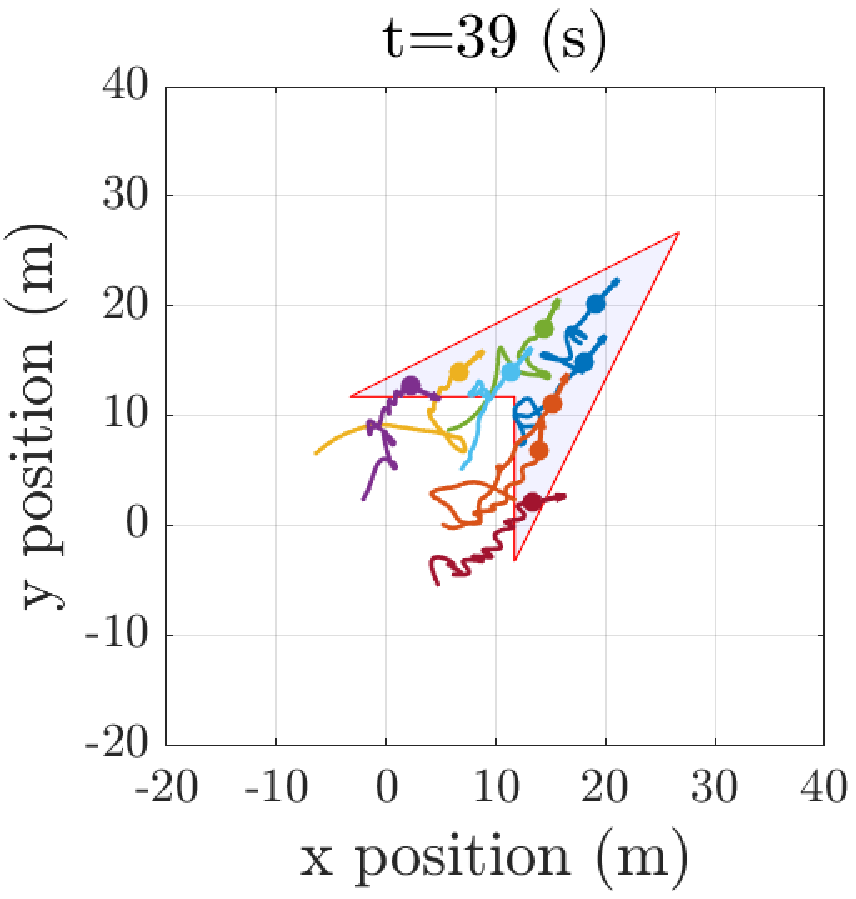}\hspace{-0.2cm}
\caption{}
\label{fig:arrow39}
\end{subfigure}
\begin{subfigure}[b]{4.2cm}
\includegraphics[width=4.2cm]{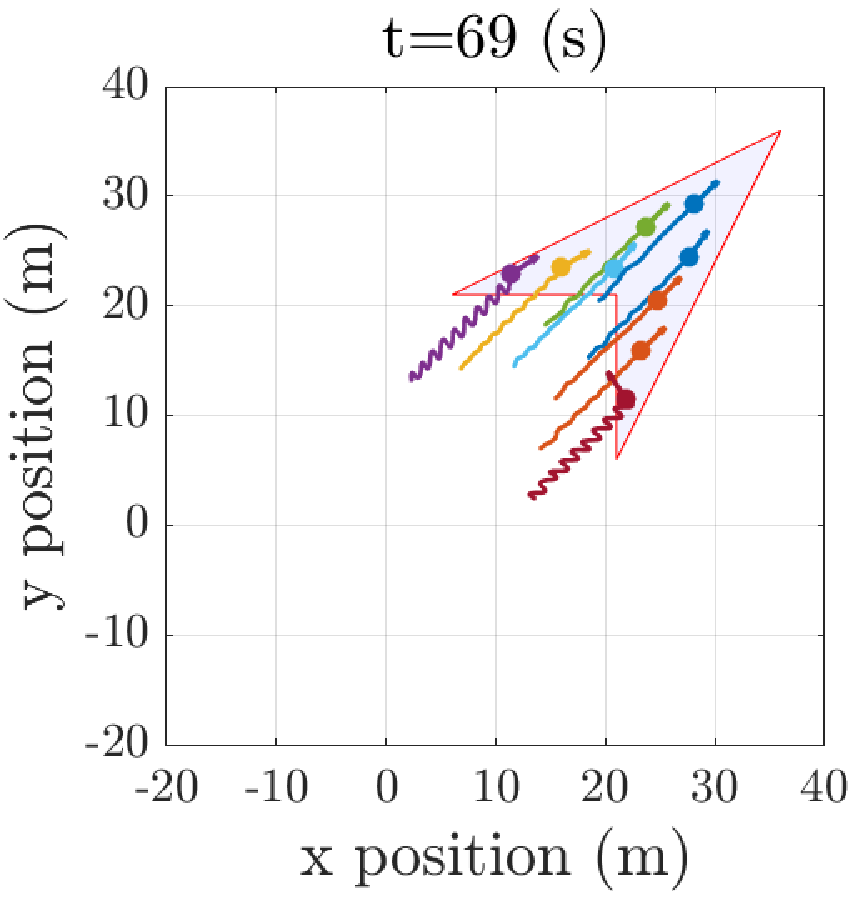}
\caption{}
\label{fig:arrow69}
\end{subfigure}
\caption{Vehicles covering and following a moving, non-convex domain, when $N=9$, $c_{r}=2\,\text{(m)}$, $v_{max}=10\,\text{(m/s)}$, $u_{max}=3 \,\text{(m/s$^{2})$}$, $t_{safety}=5\,\text{(s)}$, domain area $A=225\,\text{(m$^2$)}$ and $r_{d}=\sqrt{\frac{A}{N}}=5\,\text{(m)}$. The vehicles start in linear formation, approach and cover the domain, while following it. The vehicles lagging behind exhibit oscillations due to a bouncing effect in the narrow corners.}
\label{fig:arrow}
\end{figure}

Finally, we consider the scenario in which vehicles cover and follow a non-convex moving domain. While the domain preserves its shape, it moves with a constant velocity $v_{\Omega}=\left(0.3,0.3\right)$. Different time instants of the simulation are shown in Fig. \ref{fig:arrow}, where the tails represent the vehicle positions during the last 30 seconds of the simulation. Initially, all the 9 vehicles lie on a line perpendicular to the movement direction of the target domain, as shown in Fig. \ref{fig:arrow0}. 

We distinguish two main behaviours: during a first phase of the simulation (Figs. \ref{fig:arrow9} and \ref{fig:arrow39}) the vehicles cover the domain approximately evenly, adopting the arrow shape, while in a second phase (Fig. \ref{fig:arrow69}), a clearer domain-following behaviour is observed. The oscillations of the two vehicles that are lagging behind are the effect of their proximity to the corners. Indeed, as one of the line segments of the boundary wedge gets closer to the vehicle near the corner, it pushes it towards the other segment of the wedge, a back-and-forth motion that causes the zigzagging.

Unlike the convex case, in non-convex domains the projection on the boundary for points outside of the domain may not be unique; this is the case for instance of the green vehicle in the middle of the initial setup -- see  Fig. \ref{fig:arrow0}. Although the chance for a vehicle to lie in one of these states is extremely unlikely (the set of points where this happens has zero measure), this fact may yield ambiguity in the definition of the domain-vehicle force. 
We mitigate this issue by considering the contribution from only one of the multiple projection points; consequently, the numerical time evolution may depend on the chosen projection method.

\section{Conclusion}

In this paper, we proposed a method for safe multi-vehicle coordination which allows a swarm of vehicles to cover any compact planar shape. Vehicles are modeled using second-order dynamics with bounded acceleration, which is more realistic than the first-order models commonly used for coverage problems.
The coverage controller is based on artificial potentials among vehicles, and between each vehicle and the domain of interest. 
Using Lyapunov analysis, we prove that our algorithm guarantees coverage.
The safety controller is based on Hamilton-Jacobi reachability, which guarantees pairwise collision avoidance.
Besides drastically reducing collision count, the safety controller also helps break symmetries and lead to faster convergence to steady states.
We demonstrate our approach on three representative simulations involving a square domain, a triangle domain, and a non-convex moving domain.

Immediate future work includes parameter tuning to reduce oscillations in the vehicles' movement, studying three-dimensional coverage, investigating geometrical properties of steady states, investigating scenarios involving partial information, and implementing our approach on robotic platforms.


\bibliography{ifacconf}             
                                                   






\end{document}